\documentclass[fleqn]{aa501}
\usepackage{epsf}
\usepackage{graphicx}
\newcommand{\femi}{$f_{\rm emi}(E)$}

\newcommand{\kms}{km\,s$^{-1}$}
\newcommand{\erg}{erg\,s$^{-1}$}
\newcommand{\ergs}{erg\,cm$^{-2}$s$^{-1}$}
\newcommand{\ergcs}{erg\,cm$^{3}$s$^{-1}$}

\newcommand{\phocs}{photons\,cm$^{-2}$s$^{-1}$}

\newcommand{\gcs}{g\,cm$^{-2}$s$^{-1}$}

\newcommand{\gs}{g\,s$^{-1}$}

\newcommand{\lx}{$L_{\rm x}$}
\newcommand{\lxx}{$L_{\rm x,1}$}
\newcommand{\lacc}{$L_{\rm acc}$}
\newcommand{\laci}{$L_{\rm acc,i}$}

\newcommand{\mdotx}{${\dot M_{\rm x}}$}
\newcommand{\mdotbb}{${\dot M_{\rm bb-disk}}$}

\newcommand{\msun}{$M_{\odot}$}

\newcommand{\mwd}{$M_1$}
\newcommand{\msec}{$M_2$}
\newcommand{\rwd}{$R_1$}

\newcommand{\ri}{$r_{\rm i}$}
\newcommand{\rie}{$r_{\rm i,e}$}

\newcommand{\rco}{$r_{\rm co}$}

\newcommand{\ten}[2]{#1~10^{#2}}
\newcommand{\kts}{k$T_{\rm s}$}
\newcommand{\teff}{$T_{\rm eff}$}
\newcommand{\tmin}{$T_{\rm min}$}
\newcommand{\tmax}{$T_{\rm max}$}
\newcommand{\tirr}{$T_{\rm irr}$}

\newcommand{\pspin}{$P_{\rm spin}$}
\newcommand{\porb}{$P_{\rm orb}$}
\newcommand{\pbeat}{$P_{\rm beat}$}

\begin{document}
\title{An HST parallax of the distant cataclysmic variable V1223\,Sgr,
its system parameters, and accretion rate\thanks{Based on observations
made with the NASA/ESA Hubble Space Telescope, which is operated by
the Association of Universities for Research in Astronomy, Inc., under
NASA contract NAS 5-26555. These observations are associated with
proposal \#9230.}}

\author {K.\ Beuermann \inst{1}
     \and   Th. E. Harrison  \inst{2,**}
     \and   B. E. McArthur \inst{3}
     \and   G. F. Benedict \inst{3}
     \and   B.T.\ G\"ansicke \inst{4}}

\institute{Universit\"ats-Sternwarte G\"ottingen, Geismarlandstr.~11, 
D-37083 G\"ottingen, Germany, beuermann@uni-sw.gwdg.de
\and
New Mexico State University, Box 30001/MSC 4500, Las Cruces, NM 88003,
tharriso@nmsu.edu\thanks{Visiting Astronomer, Cerro Tololo
Inter-American Observatory, National Optical Astronomy Observatory,
which is operated by the Association of Universities for Research in
Astronomy, Inc., under cooperative agreement with the National Science
Foundation.}  
\and 
McDonald Observatory, University of Texas, Austin, TX 78712,
mca@barney.as.utexas.edu, fritz@astro.as. utexas.edu
\and
Department of Physics and Astronomy, University of Southampton,
Highfield, Southampton SO17\,1BJ, UK, btg@astro.soton.ac.uk}
\date{Received October 2, 2003 / Accepted February 10, 2004}
 
\authorrunning{K. Beuermann et al.}  
\titlerunning{{\sl HST} parallax and system
parameters of V1223\,Sgr}


\abstract{Using the {\it Hubble space Telescope Fine Guidance Sensor},
we have measured the trigonometric parallax of the bright cataclysmic
variable 1223\,Sgr. The absolute parallax is $\pi_{\rm abs}= 1.96\pm
0.18$\,mas, making V1223\,Sgr the most distant CV with a
well-determined trigonometric parallax. This distance, a Lutz-Kelker
correction, and the previously measured extinction yield an absolute
visual high-state magnitude $M_{\rm V} =4.0\pm0.2$.  We outline a
model, which is consistent with the observed spin-down of the white
dwarf and provides for much of the UV/optical emission by
reverberation of X-rays. From previous X-ray {and UV/optical}
data, we derive {an accretion luminosity \lacc\ =
$\ten{(2.6\pm0.8)}{34}$\,\erg}, a white dwarf mass $M_1 =
0.93\pm0.12$\,\msun, and an accretion rate $\dot M =
\ten{(1.4\pm0.3)}{17}$\,\gs.
 } 

\maketitle

\section{Introduction}
\label{sec-intro}

V1223\,Sgr (4U1849-31) is one of the brightest confirmed intermediate
polars, both optically and in X-rays. The only previous distance
estimate of 600\,pc was based on the \mbox{observed $E_{\rm B-V}=0.15$
and a mean reddening} of 0.25\,mag\,kpc$^{-1}$ (Bonnet-Bidaud et
al. 1982). Since V1223 Sgr is not included in ground-based parallax
programs, we have obtained a high-precision parallax using the {\it
Hubble Space Telescope Fine Guidance Sensor} ({\it FGS}). This
observation increases the number of CV parallaxes measured with the
{\it HST FGS} to eight (Harrison et al. 1999, 2000, 2003, McArthur et
al. 1999, 2001, Beuermann et al. 2003). 

V1223\,Sgr shows modulations at the orbital period \porb=3.366\,hrs, the
spin period of the white dwarf \pspin=745.5\,s (Osborne et
al. 1985), and the beat period \pbeat=794.38\,s (Steiner et al. 1981). The
UV/optical spectral energy distribution was interpreted by
Bonnet-Bidaud et al. (1982) and Mouchet (1983) in terms of an
accretion disk which reaches close to the white dwarf and should cause
it to be spun up. Unexpectedly, however, the white dwarf was found to
spin down (Jablonski \& Steiner 1987, van Amerongen et al. 1987)
suggesting that the disk stays much further away from the white
dwarf. Our accurate distance allows us to determine the luminosity of
V1223\,Sgr, estimate its accretion rate, and to show that a standard
intermediate-polar model is consistent with the available
observations.

\section{Observations, data reduction, and results}

\subsection{$HST$ observations}

\begin{table*}[t]
\caption{Photometric and spectroscopic data for the V1223 Sgr reference frame.}
\label{tab:input}
\begin{tabular}{lccccccccc}
\hline \noalign{\smallskip}

Star & $V$ & $B$-$V$ & $V$-$R$ & $V$-$I$ & $J$-$H$ & $H$-$K$ & $K$ &
$A_{\rm V}$ (mag) &Spectral Type  \\ 
\noalign{\smallskip} \hline \noalign{\smallskip} 
V1223 Sgr & 13.18& 0.02& 0.05& 0.15& 0.16& 0.06& 12.68& \hspace{1.2mm}0.47$^{1}$ &         \\
Ref-\#1   & 12.81& 1.80& 1.04& 2.24& 1.08& 0.14&  8.07& 0.81                     & M1III   \\
Ref-\#2   & 12.34& 0.59& 0.33& 0.66& 0.24& 0.11& 10.89& 0.41                     & F5V     \\
Ref-\#3   & 13.01& 1.16& 0.59& 1.13& 0.57& 0.12& 10.40& 0.40                     & G9.5III \\
Ref-\#4   & 13.04& 1.17& 0.60& 1.18& 0.70& 0.09& 10.28& 0.50                     & G9.5III \\
Ref-\#5   & 13.44& 1.40& 0.71& 1.34& 0.70& 0.15& 10.30& 0.84                     & K1III   \\
\noalign{\smallskip} \hline \noalign{\smallskip}
\end{tabular}

$^1$ From $E_{\rm B-V} =0.15$ (Bonnet-Bidaud et al. 1982).
\end{table*}

\begin{table*}[t]
\caption{Astrometric data and derived
spectroscopic parallaxes for the V1223 Sgr reference frame.}
\label{tab:input}
\begin{tabular}{l@{\hspace{3mm}}cc@{\hspace{2mm}}ccc@{\hspace{2mm}}cccrcccc}
\noalign{\smallskip} \hline \noalign{\smallskip}
Star & $\alpha_{\rm 2000}$ & \multicolumn{2}{c}{PM$_{\alpha}$ (mas/yr)} & $\delta_{\rm 2000}$ 
& \multicolumn{2}{c}{PM$_{\delta}$ (mas/yr)} &  $\xi^{~1}$   &  $\eta$ & 
$\sigma_{\xi}$ & $\sigma_{\eta}$ & $\pi$ \\
 &  & UCAC2 & FGS &  & UCAC2 & FGS & (mas) & (mas) & (mas) & (mas) & (mas)\\
\noalign{\smallskip} \hline \noalign{\smallskip}
V1223\,Sgr& 18 55 02.3 &   +2.2$\pm$2.6 &   +2.0$\pm$0.2 & $-$31 09 49.5 &\hspace{-1mm}$-$23.7$\pm$2.9 & $-$24.6$\pm$0.2 &\hspace{-2mm}$-$67.090 & 790.152 & 0.37 & 0.35 &      \\
Ref-\#1   & 18 54 53.3 & $-$8.1$\pm$3.8 & $-$9.2$\pm$0.2 & $-$31 10 21.4 & $-$ 8.1$\pm$4.4 & $-$ 5.8$\pm$0.2 &    50.571 & 809.254 & 0.45 & 0.36                         & 0.33 \\
Ref-\#2   & 18 54 42.3 & $-$0.2$\pm$2.6 & $-$0.5$\pm$0.2 & $-$31 09 10.1 &\hspace{-1mm}$-$13.7$\pm$3.0 & $-$13.9$\pm$0.2 &\hspace{-2mm}  183.174 & 722.876 & 0.60 & 0.49 & 1.91 \\
Ref-\#3   & 18 54 49.2 & $-$7.9$\pm$3.8 & $-$5.5$\pm$0.2 & $-$31 07 59.6 &\hspace{-1mm}$-$18.0$\pm$4.4 & $-$17.8$\pm$0.2 &    87.710 & 662.492 & 0.42 & 0.45             & 0.40 \\
Ref-\#4   & 18 55 01.5 &   +1.3$\pm$3.8 & $-$0.7$\pm$0.2 & $-$31 07 53.2 & $-$ 7.4$\pm$4.4 & $-$ 7.4$\pm$0.2 &\hspace{-2mm}$-$70.136 & 673.396 & 0.41 & 0.45             & 0.41 \\
Ref-\#5   & 18 55 07.1 &   +3.3$\pm$3.8 &   +4.4$\pm$0.2 & $-$31 09 55.7 & $-$ 2.4$\pm$4.4 & $-$ 0.9$\pm$0.2 &\hspace{-3mm}$-$128.303 & 803.079 & 0.49 & 0.54            & 0.37 \\  
\noalign{\smallskip} \hline \noalign{\smallskip}
\end{tabular}

$^1$ $\xi$ and $\eta$ are the relative positions in arcseconds from
Eqs.~(3) and (4), $\sigma_{\xi}$ and $\sigma_{\eta}$ their standard
deviations.
\end{table*}

The process for deriving a parallax for a cataclysmic variable from
{\it HST FGS} observations has been described in papers by McArthur et
al. (2001, 1999) and Harrison et al. (1999).  The process used here is
nearly identical to those efforts, as well as our recent program on EX
Hya (Beuermann et al. 2003). An {\it FGS} program consists of a
series of observations of the target of interest, and a set of four or
more reference stars located close to that target.  Typically, three
epochs of observations, each comprised of two or more individual {\it
HST} pointings (orbits), are used to solve for the variables in the
series of equations that define a parallax solution. For V1223 Sgr, we
obtained observations on four different epochs (2000 September, 2001
March, 2001 September, and 2002 September) during which nine orbits of
{\it HST} time were used. The extra epoch of observation was essential
in deriving the high precision in the smallest parallax we have yet
measured using the {\it FGS}. Extensive calibration data, as well as
estimates of the distances and proper motions of the reference stars,
are required to obtain a robust parallax solution.

\subsection{Spectroscopic parallaxes of the reference frame}

We have used a combination of spectroscopy and photometry to estimate
spectroscopic parallaxes for the reference stars. 
The optical \mbox{$BVRI$} photometry of the five reference stars was
obtained on 2001 March 13 using the CTIO 0.9\,m telescope and the
Cassegrain Focus CCD
imager\footnote{http://www.ctio.noao.edu/cfccd/cfccd.html}. These data
were calibrated to the standard system using observations of Landolt
standards. The final photometric data set is listed in
Table~1. Included is the Two Micron All-Sky Survey (2MASS) $JHK$
photometry of the reference stars, transformed to the homogenized
system of Bessell \& Brett (1988) using the transformation equations
from Carpenter (2001). Typical error bars on the photometry are $\pm$
0.02 mag for the $V$-band measurements, and $\pm$ 0.03 mag for the
optical colors. The 2MASS photometry for the five reference stars has
error bars of $\pm$ 0.02 mag.
Optical spectroscopy of the reference stars, and a number of MK
spectral type templates, was obtained on 2001 March 9 and 10 using the
CTIO 1.5\,m telescope with the Cassegrain
Spectrograph\footnote{http://www.ctio.noao.edu/spectrographs/60spec/manual/}
with the 831 {\it l}/mm \mbox{``G-47''} grating and a two arcsecond
slit, giving a spectral resolution of 0.56 \AA/pix. We estimated
spectral types of the reference stars from a comparison of the spectra
with those of the MK-templates and present the results in the last
column of Table~1. 
The visual extinctions of the reference stars (column\,9 of Table~1)
were estimated from a comparison of their spectral types and the
photometric data, {using the standard relations from Reike \&
Lebofsky (1985).}  Finally, spectroscopic parallaxes were derived
using the spectral types and visual extinctions. The results are
listed in the last column of Table~2. {To determine these values,
we used the {\it Hipparcos} calibration of the absolute magnitude for
main sequence stars by Houk et al. (1997), and that for giant stars
tabulated by Drilling \& Landolt (2000). The distinction between main
sequence stars and giants is facilitated by (i) the dichotomy of the
$M_V$-distribution in the Hertzsprung-Russell diagram for stars of
spectral type G and later and (ii) the ability of the Bayesian-like
astrometric software to identify an object with a true parallax
substantially different from the one used as the starting value in the
iterative approach.} Details of the procedure are given in our
previous papers cited above.  For the astrometric solution discussed
below, we assumed error bars of $\pm$ 20\% on our spectroscopic
parallaxes.

The reference stars span a wide range in spectral type, and have
significant extinctions.  Four of the five reference stars turn out to
be late-type giants, providing the exceptionally quiet reference
frame, which is an essential for the high quality of the parallax
derived for V1223 Sgr. Three of them (\#2, \#3, and \#4) have
extinctions similar to that of V1223\,Sgr which has $A_{\rm V}=0.47\pm
0.02$ (Bonnet-Bidaud et al. 1982). {It is somewhat disquieting
that $A_V$ does not correlate better with~$\pi$. A similar
situation, however, was also found in the fields of RR\,Lyr (Benedict et
al. 2002a) and $\delta$\,Cep (Benedict et al. 2002b).}

\subsection{The astrometric solution}

The data reduction process for deriving a parallax {of the target} from
FGS observations was identical to previous efforts, except for the
fact that the astrometer has changed from FGS3 to FGS1R, successfully
calibrated by Benedict et al. (2001).  With the positions measured by
FGS1R, we determine the scale, rotation, and offset plate constants
for each observation set relative to an arbitrarily adopted constraint
plate.  The solved equations are
\begin{eqnarray}
x' & = & x + lc(B-V)\\
y' & = & y + lc(B-V)\\
\xi & = & Ax' + By' + C + R_{\rm x}(x'^2 + y'^2) - \mu_x t_{\rm rel} - 
P_{\alpha}\pi_x\\
\eta & = & Dx' + Ey' + F + R_{\rm y}(x'^2 + y'^2) - \mu_y  t_{\rm rel} - 
P_{\delta}\pi_y
\end{eqnarray}
where $\it x$ and $\it y$ are the rectangular HST coordinates and $\it
lc$ is the derived lateral color correction for color $B-V$ (Benedict
et al. 1999). $A, B, D$ and $E$ are a set of scale plate constants,
$C$ and $F$ are zero point corrections, $R_{\rm x}$ and $R_{\rm y}$
are radial terms, $\mu_x$ and $\mu_y$ are the proper motions, $t_{\rm
rel}$ is the time of the observation minus the constraint plate time,
$P_{\alpha}$ and $P_{\delta}$ are parallax factors, and $\it \pi_x$
and $\it \pi_y$ are the parallaxes in $x$ and $y$. 

\begin{figure}[t]
\includegraphics[width=4.3cm]{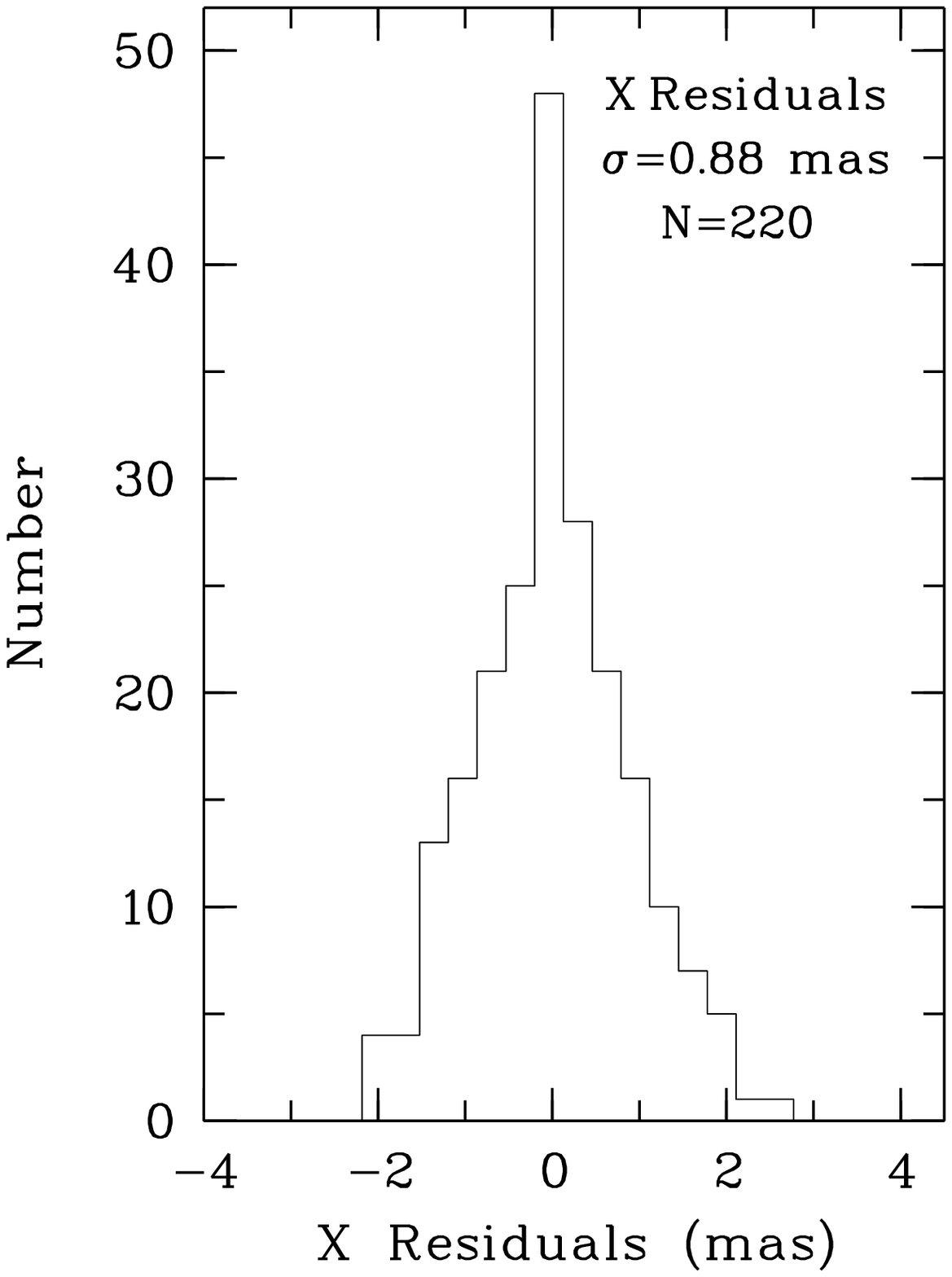}
\hfill
\includegraphics[width=4.3cm]{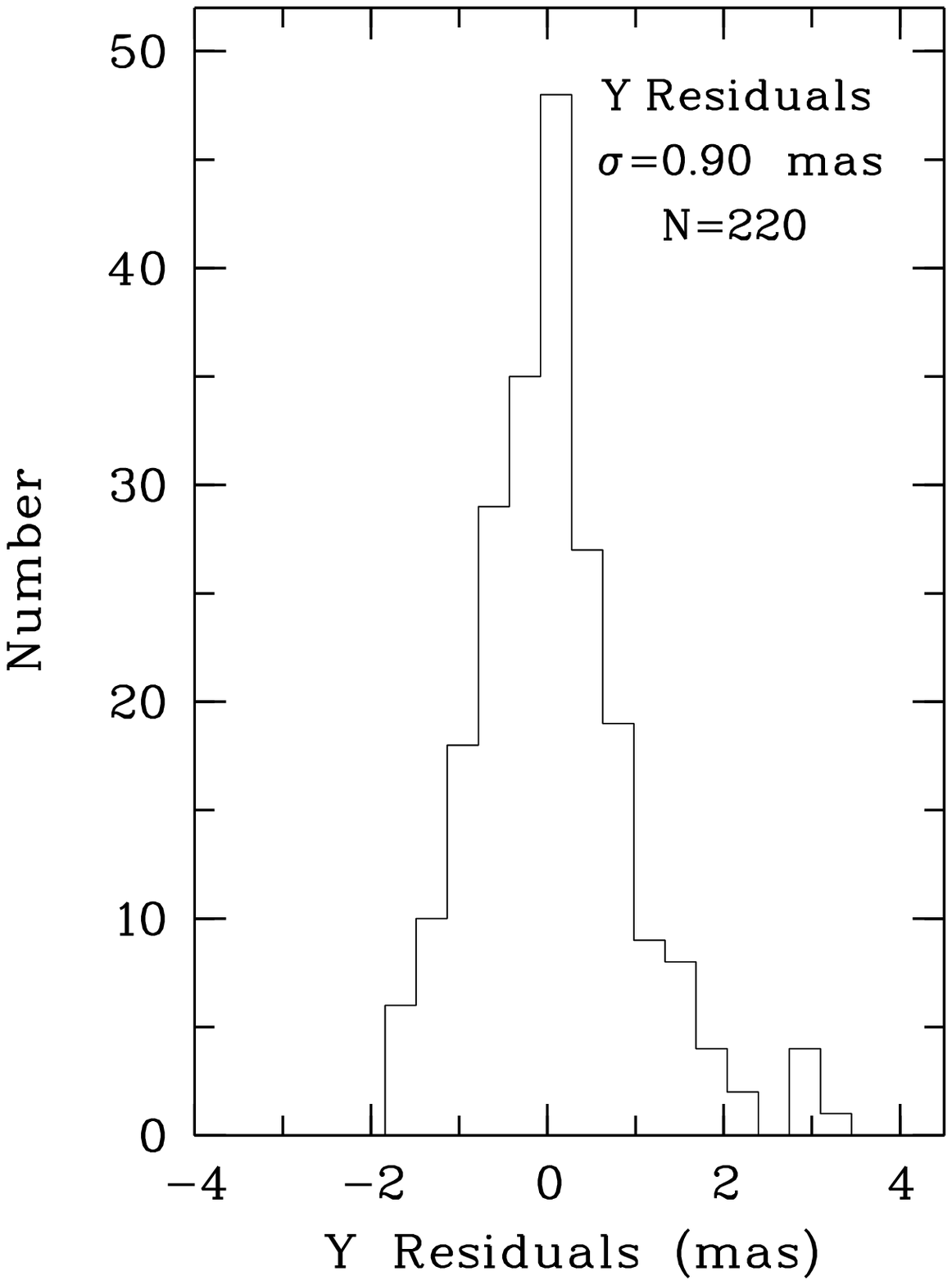}
\caption[ ]{Histograms of the x and y residuals obtained from modeling
V1223 Sgr and its reference frame. }
\label{fig-1}
\end{figure}

As in all our previous astrometric analyses, we employed GaussFit
(Jefferys, Fitzpatrick, \& McArthur, 1987) to minimize $\chi^2$.
Since we had five high-quality reference star observations at each
epoch, we were able to use the eight-parameter model, {instead of the
six-parameter model} used in our previous parallax measurements,
{which had $A = E$ and $B = -D$.} The additional parameters yielded a
significantly improved $\chi^2$ compared with the simpler model. {The
input proper motions for V1223\,Sgr and the reference stars were taken
from the USNO CCD Astrograph Catalog, the ``UCAC2'' (Zacharias et al.
2004)\footnote{ftp://cdarc.u-strasbg.fr/cgi-bin/VizieR, catalogue
I/289}, and are listed in columns 3 and 6 of Table~2. The
Bayesian-like reduction software models the proper motions and the
spectroscopically determined parallaxes of the reference stars as
observations with errors to produce an absolute (not relative)
parallax for V1223\,Sgr.  It yields the final proper motions listed in
columns 4 and 7 of Table~2. The internal consistency of the solution
obtained for the V1223\,Sgr field is very good. Fig.~\ref{fig-1} shows
the histograms of the residuals for all 220 individual $x$ and $y$
positional measurements of the target and the reference
stars. Gaussian fits give standard deviations of 0.88 and 0.90\,mas,
respectively, which is better than usually obtained for parallax
measurements with the {\it HST FGS}. Columns 8 to 11 of Table~2 list
the ``catalog'' positions $\xi$ and $\eta$ and their standard
deviations $\sigma_{\xi}$ and $\sigma_{\eta}$ which result from the
fit of Eqs. (3) and (4) to the time series of the individual
observations. The frequency-weighted mean values of these standard
deviations are $\sigma_{\xi} = 0.43$\,mas and
$\sigma_{\eta}=0.46$\,mas.  Reference star \#3 has standard deviations
in both coordinates as low as the other stars, confirming that its
input parallax is not grossly in error. Hence, we must presently
accept that it is a giant at a distance of $\sim 2.5$\,kpc, although
its implied tangential velocity is large.}

The final {absolute} parallax of V1223 Sgr is {$\pi$ = 1.96 $\pm$ 0.18
mas.}  The fine precision of this result is directly attributed to the
very stable reference frame afforded by the distant giants. The
limiting factor in the error budget turns out to be the remaining
uncertainty in the proper motions of the reference frame.

\subsection{The distance to V1223 Sgr}

The measured distance to V1223\,Sgr is $d = 1/\pi =
510^{+52}_{-43}$\,pc. As pointed out by Lutz \& Kelker (1973),
parallaxes suffer from a systematic error in addition to the
observational one in the sense that the most probable true parallax
is smaller than the observed absolute parallax because the number of
objects per $\pi$-interval increases as $\pi^{\rm -n}$ with $n=4$ for
a constant space density. More detailed analyses of the problem
including the magnitude and space density distributions of the parent
population have been considered, e.g., by Hanson (1979), by Smith
(1987), and by Oudmaijer et al. (1998). Here, we have simply assumed
that the parent population has an absolute magnitude at the brighter
end of the range 0 $<$ M$_{\rm V}$ $<$ 10, giving an index of $n=3.0$
for the space density (Hanson 1979). The Lutz-Kelker correction {is
usually quoted as a correction in visual magnitude} and, for this
choice of parameters, is $\Delta M_{\rm V} = -0.071$. {The most
probable} distance modulus of V1223\,Sgr including the Lutz-Kelker
correction is $m-M=8.61^{+0.21}_{-0.19}$ and the most probable
distance including this correction is $d=527^{+54}_{-43}$\,pc. This is
the largest distance of a CV so far measured with the {\it HST FGS}.

\section{System parameters and accretion rate of V1223~Sgr}

\subsection{Absolute magnitude}

In its normal high state, V1223\,Sgr has a long-term mean magnitude of
$B = 13.1$ (Garnavich \& Szkody 1988), around which it varies by $\pm
0.5$\,mag. Of this, the orbital and spin modulation accounts for about
$\pm 0.3$\,mag. There are also long-term trends and a time span
between 1937 and 1951 when it experienced drops to much fainter states
(Garnavich \& Szkody 1988). With $B-V\simeq 0$ in the high state
(Table~2), $V$ and $B$ are practically the same. From the depth of the
2200\AA\ feature, Bonnet-Bidaud et al. (1982) derived $E_{\rm
B-V}=0.15$ with an error of about 0.01, which yields a visual absorption
$A_{\rm V} = 0.47\pm 0.03$. This value of the absorption, the mean
visual magnitude of $V=13.1$, and our new distance modulus give an
absolute mean visual magnitude in the normal high state of $M_{\rm V}
= 4.0\pm 0.2$. This value has still to be corrected for inclination
effects.

\subsection{Secondary star and inclination}

{Penning (1985) and Watts et al. (1985) assumed that the mass of the
secondary in V1223\,Sgr is that of a Roche-lobe filling main sequence
star, $M_2 = 0.40$\,\msun. In many CVs, however, the secondaries seem
to be expanded over their main sequence radii and have smaller masses
than the main-sequence assumption suggests (e.g. Beuermann et
al. 1998). We adopt $M_2 = 0.40$\,\msun\ as the nominal secondary mass
and comment on the effect a mass as low as 0.25\,\msun\ will have on
the the results.  For a spectral class M4, a secondary in this mass
range has a $K$-band magnitude of $K=15.1-15.4$ and contributes
8--11\% of the observed $K$-band flux (Table~1).}

{According to Watts et al. (1985),} the radial velocity amplitude
of the orbital motion of the white dwarf as derived from the wings of
the Balmer emission lines is $K_1 = 56\pm13$\,\kms. {Penning
(1985), using a superior time resolution, found that the Balmer and
HeII$\lambda4686$ radial velocities were modulated, in addition, at
the spin period (see his Figs.\,9 and 3 and note that the spin
modulation is clearly present in his Fig.\,3). Penning's $K_1$
agrees closely with that of Watts et al., which we adopt here. As
noted already by these authors, the implied inclination is small. For
$M_1 = 0.93$\,\msun\ (see below), we obtain $i = (24\pm7)^{\circ}$
with $M_2=0.40$\,\msun\ and $i=(37\pm10)^\circ$ with
$M_2=0.25$\,\msun.}

\subsection{X-ray and XUV emission of V1223\,Sgr}

\subsubsection{The X-ray spectral analysis of Beardmore et al.}

{Our} derivation of the white dwarf mass \mwd\ and the
accretion rate $\dot M$ rests on the results of the X-ray spectral
analysis of Beardmore et al. (2000). These authors fitted a
multi-temperature shock model to the {\it ASCA} and {\it Ginga} X-ray
spectra of V1223\,Sgr, which included a treatment of the X-albedo and
of internal absorption within the source. From this fit, they deduced
a shock temperature of \kts\ = $43^{+13}_{-12}$\,keV and the emitted
spectrum \femi\ as a function of photon energy~$E$ (called 'incident'
spectrum by them). Integration of \femi\ over all photon energies
above the Lyman edge gives the total flux $F_{\rm
x}=3.30\,10^{-10}$\,\ergs\ (Table~3, line 1), which we call 'X-ray
flux' because only 3\% of {it} is emitted below 0.1\,keV. $F_{\rm
x}$ refers to \kts\,=\,43\,keV and scales, to a first approximation,
as \kts. {The X-ray luminosity \lx\ then scales as \kts, too. The
emitted spectrum \femi\ is based on the observed {\it ASCA} spectrum
for photon energies $E>0.4$\,keV, but for still lower energies it
merely represents an extrapolation of the multi-temperature thermal
model spectrum. We demonstrate in the next Section, however, that a
substantial additional XUV source does not exist in V1223\,Sgr.

The observed X-ray luminosity based on the Beardmore et al. (2000)
emitted spectrum and our Lutz-Kelker corrected distance from
Sect. 2.4, is \lxx$=4\pi d^2\,F_{\rm
x}=\ten{(1.11\pm0.36)}{34}$\,\erg, where the error accounts for the
uncertainties in the normalization of the observed X-ray flux, in
\kts, and in $d$.  We assume that \lxx\ represents the luminosity of
one accreting pole and that the second pole is hidden behind the white
dwarf. Such a geometry is generally expected if the shock height $h$
is small compared with the white dwarf radius \rwd, as it is in the
model of Beardmore et al., and still holds for a finite shock
height if the inclination is small and the magnetic axis is more or
less aligned with the rotational axis. We account for the second pole
by writing $L_{\rm x}=(1+\beta)\,L_{\rm x,1}$ and use $\beta=1$,
below.}

\begin{table}[t]
\caption{Integrated observed and inferred fluxes for individual
wavelength ranges (see text).}
\label{tab:obsflux}
\begin{tabular}{ll@{\hspace{5mm}}c}
\noalign{\smallskip} \hline \noalign{\smallskip}
Band            & Component                     &  Flux               \\
                &                     & (\ergs)             \\
\noalign{\smallskip} \hline \noalign{\smallskip}
0.0136--100 keV & total emitted X-ray flux$^{1}$&\hspace{3mm}$\ten{3.30}{-10}$\\
$>54$\,eV       & additional XUV source         &\hspace{6.5mm}$<10^{-11}$ \\
$13.6-54$\,eV   & not observable                & \hspace{4mm}?\\
$912-1250$\,\AA & unobserved FUV$^2$            & $\ten{(2-3)}{-10}$\\
$>1250$\,\AA    & observed UV/optical$^{3}$/IR$^{4}$& \hspace{3mm}$\ten{7.59}{-10}$\\
\noalign{\smallskip} \hline \noalign{\smallskip}
\end{tabular}

$^1$ Integral of 'incident' spectrum of  Beardmore et al. (2000);
$^2$ extrapolated from $\lambda>1250$; $^3$ total mean flux from
Mouchet (1983); $^4$ this work.
\end{table}

\subsubsection{The quest for additional XUV emission}

In magnetic CVs of the polar subtype, a large fraction of the
accretion energy is released as soft X-rays and dominates the total
X-ray flux, which is used to estimate their accretion rates (Beuermann
\& Burwitz 1995). Does a soft X-ray component exist also in the
intermediate polar V1223\,Sgr?

The {XUV} luminosity can be estimated by the Zanstra method using
the observed line flux $F_{4686}$ of HeII$\lambda 4686$ photons
 with h$\nu=2.65$\,eV in \ergs. The number flux of ionizing photons needed to
produce this line is
\begin{equation}
N_{\rm >54} = F_{4686}\frac{\alpha_{\rm B}}{\varepsilon_{4686}} \frac{4\pi}{\Omega} 
\end{equation}
where $\Omega$ is the solid angle subtended by the photoionized gas,
{$\alpha_{\rm B}$ is the total recombination coefficient of He\,II
for the Baker \& Menzel Case B in cm$^3$s$^{-1}$, and $\varepsilon_{4686}$
the emissivity in the $\lambda4686$ line in \ergcs.  For the pure
recombination case, $\varepsilon_{4686}$/h$\nu_{4686}$ is the
effective recombination coefficient for this transition. We use the
results of Storey \& Hummer
(1995)\footnote{ftp://cdarc.u-strasbg.fr/cgi-bin/VizieR, catalogue
VI/64}, which take full account of collisional processes and are
tabulated for electron number densities of $N_{\rm e}=10^2 -
10^{14}$\,cm$^{-3}$ and a wide range of electron temperatures. They
yield $\varepsilon_{4686}/({\rm h}\nu_{4686}\,\alpha_{\rm B})$ =
0.216\,(0.140), 0.157\,(0.125), and 0.210\,(0.156) for $N_{\rm
e}=10^6$, $10^{10}$, and $10^{14}$\,cm$^{-3}$ and electron
temperatures $T_{\rm e}=10^4$ ($\ten{5}{4}$)\,K, respectively. The
particle density in the photoionized region of V1223\,Sgr may range
from \mbox{$>10^{14}$\,cm$^{-3}$} in the disk and in the densest parts
of the magnetically confined accretion stream to much lower values
elsewhere in the magnetosphere or in the corona of the disk and its
bulge, while the temperature is likely between $10^4$ and a few times
$10^4$\,K. The above numbers demonstrate that
$\varepsilon_{4686}/({\rm h}\nu_{4686}\,\alpha_{\rm B})=0.18\pm0.05$
is appropriate for a wide range of parameters in the photoionized
regions in V1223 Sgr.}

Steiner et al. (1981) quote an energy flux of the $\lambda4686$ line
in V1223\,Sgr of $\ten{9}{-14}$\,\ergs\ and other papers suggest
values within a factor of two {in both directions} (Bonnet-Bidaud
et al. 1982, Penning 1985, Watts et al. 1985). We adopt the Steiner et
al. value which becomes $F_{4686}=\ten{1.6}{-13}$\,\ergs\ after
correction for reddening. Because of the large internal absorption at
{soft X-ray energies, we use} $\Omega \simeq 4\pi$. {This
choice of parameters yields $N_{\rm >54}\simeq0.21$\,\phocs, with an
uncertainty of about a factor of two from the error in $F_{4686}$ and
the uncertainties in $\Omega$ and $\alpha^{\rm
eff}_{4686}/\alpha_{\rm B}$.}  For comparison, the Beardmore et
al. X-ray spectrum $f_{\rm emi}(E)$ provides 0.11\,\phocs\ with
energies above 54\,eV, of which 50\% are below 0.3\,keV. As argued
above, this number probably refers only to the pole facing the
observer. The number from both poles would then be 0.22\,\phocs,
in agreement with the number needed to produce the observed
$\lambda4686$ flux. We conclude that there is no evidence for an
additional component of XUV photons. The conservative limit to the
energy flux of a potential XUV source given in line 2 of Table~3
refers to a source which provides the same number of ionizing photons
as the X-ray source.
 
\subsection{An intermediate polar model for V1223\,Sgr}

{For further discussion of V1223\,Sgr, we adopt a simple
intermediate polar model in which an accretion disk is broken up by
the magnetic field of the primary at an inner radius $r_{\rm i}>R_{\rm
1}$. At \ri, matter couples onto the field and falls along the field
lines towards both magnetic poles of the white dwarf. We assume that
there is no viscous energy release inside \ri\ and that all
gravitationally released energy is radiated in the flow behind the
strong shock which forms near the white dwarf surface. This is the
usual assumption, which allows to treat the guided motion as a free
fall. It neglects the fact that the ionized matter in the funnel has
to get rid of the particle momenta perpendicular to the field and does
so by radiative losses. Nevertheless, photoabsorption of X-rays in the
guided stream and reemission of the absorbed energy is in line with
the model and may cause the funnel to be an intense source of
UV/optical radiation.
Outside \ri, we assume a circularly symmetric steady state disk in
which the individual annuli radiate at local effective temperatures
\teff(r) as given by Shakura \& Sunyaev (1973). In addition, we
include an inner ring at \ri, which mimics the boundary layer and
emits the energy released when the matter is braked from the Kepler
velocity to the angular velocity of the white dwarf. Following
Bonnet-Bidaud et al. (1982) and Mouchet (1983), we use blackbody
spectra, which has the advantage of simplicity, but the disadvantage
of systematically distorting the spectrum, especially at short
wavelengths.
For the outer disk, we allow for the possibility of heating by
radiation from the central source or from the elevated magnetically
guided flow by letting \teff(r) not drop below a temperature
\tmin. This is obviously a very rough approach, which we justify by
its simplicity.

The location of the inner edge of the disk is restricted by two
requirements. As a first condition, \ri\ should not exceed the
circularization radius $r_{\rm circ}$ of matter with the specific
angular momentum it carries over from the inner Lagrangian point. No
disk can form for \ri~$>r_{\rm circ}$. For $M_1 = 0.93-1.18$\,\msun\
and $M_2=0.25-0.40$\,\msun, $r_{\rm circ}$ falls in the range of
$\ten{1.06}{10}$ to $\ten{1.48}{10}$\,cm. The second condition results
from the observation that the white dwarf in V1223\,Sgr is being spun
down, which indicates that it should rotate close to equilibrium. The
equilibrium inner radius of the disk \rie\ is located inside the
corotation radius \rco~=~$(GM_1 P_{\rm
s}^2/4\pi^2)^{1/3}=\ten{1.23}{10}\,M_1^{1/3}$\,cm, with $P_{\rm
s}=745.5$\,s the spin period of the white dwarf in V1223\,Sgr (Osborne
et al. 1985). The theory of Ghosh \& Lamb (1979) with later
corrections by Wang (1987) predicts \rie$~\simeq$~0.98\,\rco. This
does not account, however, for the synchronization torque between
primary and secondary star. Warner (1996) estimates this torque from
considerations of the magnetism of the secondary star. For
conciseness, we refer to his Eqs. (22), (23), and (24), and note that
the magnitude of the effect depends not only on the uncertain magnetic
moment of the secondary star, but also on the ratio between inner edge
and spherical Alfv\'{e}n radius, usually taken to be $r_{\rm i}/r_{\rm
\mu,sph}\simeq 0.52$, and on the poorly known screening of the
magnetic moment of the primary by the accretion disk. In our preferred
model\,B (see below), the disk in V1223\,Sgr has a rather large
central hole and may be quite flimsy. We assume, therefore, that
screening is rather weak ($\delta \simeq 0.2$ in Warner's
terminology).  For the parameters of V1223\,Sgr, we then obtain
\rie$~\simeq~0.83$\,\rco~=~$\ten{1.0}{10}$\,cm, which is smaller than
$r_{\rm circ}$ for the entire range of possible values of \mwd\ and
\msec.

This theory also allows to estimate the magnetic moment of the primary
and its surface field strength.  We use \ri $\simeq 0.52\,r_{\rm
\mu,sph}$ with $r_{\rm \mu,sph} = 2^{-3/7}(GM_{\rm 1})^{-1/7}\dot
M^{-2/7}\mu^{4/7}$\,cm, where $\mu = B_{\rm 1}\,R_{\rm 1}^3$ is the
magnetic moment, $B_{\rm 1}$ is the surface field strength of the
white dwarf in the orbital plane, and the dipolar magnetic axis is
assumed to be aligned with the rotational axis. In the aligned case,
the polar field strength is $B_{\rm p}=2\,B_1$, while for an inclined
magnetic rotator, the factor between $B_{\rm p}$ and $B_{\rm 1}$ is
smaller than two.}

\subsection{Derivation of  \mwd, \mdotx, and \lx}

The free fall of matter from \ri\ to \,\rwd\ releases an accretion
luminosity \laci\ in a shock {with} temperature \kts\
\begin{eqnarray}
L_{\rm acc,i} & = & \dot M\cdot GM_{\rm 1}(1/R_{\rm 1}-1/r_{\rm i})\\
{\rm k}T_{\rm s} & = & (3\mu m_{\rm u}/8)\cdot GM_{\rm 1}(1/R_{\rm
1}-1/r_{\rm i}),
\end{eqnarray}
where $\mu = 0.617$ is the molecular weight of all particles for solar
composition, $m_{\rm u}$ is the unit mass, {and prompt
equilibration between electrons and ions in the shock is assumed.
\mbox{X-ray} emission is the major channel of energy release in the
post-shock region, but cyclotron emission may also contribute
despite the negative results of polarization measurements (Cropper
1986). We represent the cyclotron fraction by a parameter $\alpha$ and
write
\begin{equation}
L_{\rm acc,i} = (1+\alpha)\,L_{\rm x} = (1+\alpha)(1+\beta)\,L_{\rm x,1}.
\end{equation}
The quantity $\alpha$, which depends on the field strength and particle density in
the post-shock region (e.g. Woelk \& Beuermann 1996), is of the
order of 0.1.}

Eq.\,(7) can be solved for \mwd\ with \kts\ = $43^{+13}_{-12}$\,keV
(Beardmore et al. 2000), an adopted value for \ri\ and a mass radius
relation of white dwarfs \rwd(\mwd). We use the relation \mbox{$R_{\rm
1} = \left[1.463-0.885\,(M_{\rm 1}/M_\odot)\right]\cdot10^9$\,cm},
which we found from fitting the radii of CO white dwarf models by Wood
(1995) with a thick hydrogen envelope, an effective temperature of
$\ten{3}{4}$\,K, and masses between 0.80 and 1.20\,\msun.  We use two
values of \ri, which correspond to the two models considered below,
$r_{\rm i}=\ten{8.0}{8}$\,cm and $r_{\rm i}=\ten{1.0}{10}$\,cm (see
Sects. 3.6.1 and 3.6.2). The white dwarf masses for these two cases
are \mwd=$1.18\pm0.06$\,\msun\ and \mwd=$0.93\pm0.12$\,\msun,
respectively. A white dwarf mass lower than $\sim0.8$\,\msun\ can be
excluded, as noted already by Beardmore et al. (2000).

{We derive the accretion rate \mdotx\ needed to generate the
emitted X-ray spectrum \femi\ by eliminating G\mwd($1/$\rwd $-1/$\ri)
from Eq.\,(6) with help of Eq.\,(7) and replacing \laci\ from
Eq.\,(8). With \lxx\ and $\beta=1$ from Sect. 3.3.1, we then obtain
\mbox{$\dot M_{\rm x} = (1.24\pm0.24)\,10^{17}(1+\alpha)$}~\gs, where
the error depends primarily on the remaining uncertainty in $d$ and
the value of $\alpha$. Note that \mdotx\ is independent of the white
dwarf mass. The X-ray luminosity is $L_{\rm x} = 2\,L_{\rm x,1} =
(2.23\pm0.71)\,10^{34}$~\erg}.

\subsection{Interpretation of the overall UVptical/IR spectral 
energy distribution}

{In what follows, we interprete the observed UV/optical spectral
energy distribution in terms of two simple models and list their
parameters in Table~4. Model~A is that of a luminous accretion disk,
which has its inner edge close to the white dwarf (Bonnet-Bidaud et
al. 1982, Mouchet 1983). Model B has the inner edge at $r_{\rm
i}=\ten{1.0}{10}$\,cm and requires a different origin for most of the
UV emission.
We use a distance of 527\,pc and a radial velocity amplitude of the
white dwarf $K_1=56$\,\kms\ (Section 3.2). We quote the results in
Table\,4 for the nominal secondary mass $M_2=0.40$\,\msun\ and comment
on the effect of choosing a lower value of $M_2$.} 

Figure\,2 shows the mean orbital and rotational spectral energy
distribution of V1223\,Sgr in its normal bright mode. It is based on
the IUE spectrophotometry and optical photometry of Bonnet-Bidaud et
al. (1982) and Mouchet (1983) (solid circles) and is supplemented by
the (non-simultaneous) 2MASS infrared photometry from Table~1 (open
circles). The integrated flux for the wavelength range $\lambda >
1250$\,\AA\ is given in Table~3, line\,5, and an estimate of the flux
between the Lyman edge and 1250\AA\ in line 4.

\subsubsection{Model A: steady state accretion disk}

{The model of a luminous accretion disk (Bonnet-Bidaud et
al. 1982, Mouchet 1983) can account for the observed
UV/optical/infrared emission. The model} requires a high white dwarf
mass, an outer {disk} radius of 90\% of the Roche radius, an inner
radius close to the white dwarf, and an accretion rate \mdotbb\
$=\ten{3.1}{17}$\,\gs, which exceeds \mdotx\ from above. The best fit
to the spectral flux distribution is shown in Fig.\,2 (upper data set
and curve) and the parameters are listed in Table~4. The spacing
between \ri\ and \rwd\ is tailored to yield \kts\ = 43\,keV.
{The derived accretion rate $\dot M_{\rm bb-disk}$ represents an
overestimate because the blackbody disk and its boundary layer at
\ri\ emit 2/3 of the total flux in the Lyman continuum, while typical
CV disks have the Lyman jump strongly in absorption.  The blackbody
disk is a coarse approximation and we quote \mdotbb\ in Table~4 in
brackets to indicate this uncertainty.}
{If the secondary mass is reduced from 0.40 to 0.25\,\msun, the
blackbody accretion rate increases to \mdotbb\ $=\ten{5.0}{17}$\,\gs,
partly because the now larger inclination of $44^\circ$ implies a
smaller projected area of the disk. }

\begin{figure}[t]
\includegraphics[width=8.8cm]{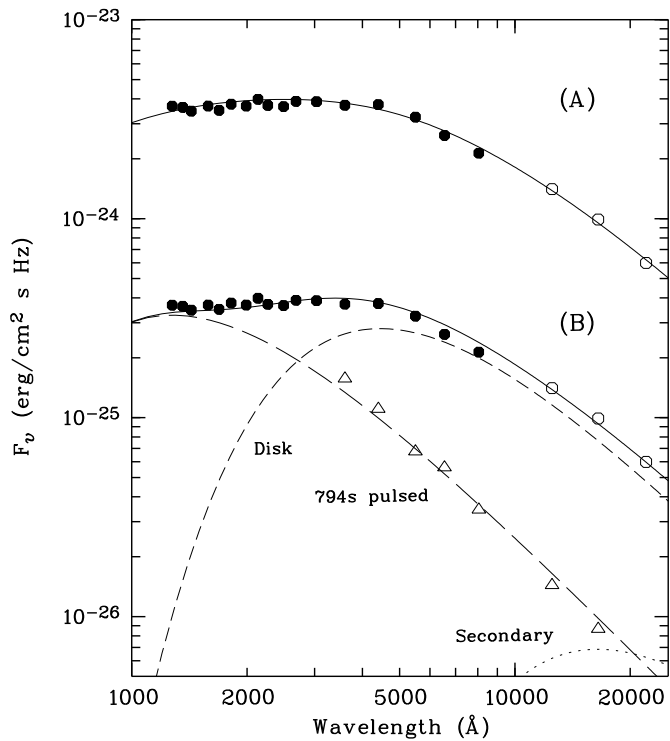}
\caption[ ]{Overall energy distribution of V1223 Sgr for $\lambda >
1000$\,\AA\ (see text). The data are shown twice with two different
model fits. Model~A (top data set, fluxes multiplied by 10) is for a
luminous steady state accretion disk and model~B (bottom data set,
fluxes as observed) is for a truncated disk plus reprocessed
component. }
\end{figure}

The real problem of the disk model is the location of the inner edge
of the disk far inside the corotation radius, which is incompatible
with the observed spin down of the white dwarf (van Amerongen et
al. 1987, Jablonski \& Steiner 1987). {The theory presented above
suggests that \ri\ should be about an order of magnitude larger than
assumed in this model.} Whatever the remaining uncertainties in the
theory, the inner edge of the disk can not be close to the white
dwarf.

\begin{table}[t]
\caption{Parameters of disk models A and B. For both \mbox{models}, we
assume $M_2=0.40$\,\msun, $K_1=56$\,\kms, and $d = 527$\,pc (see text).}
\label{tab:models}
\begin{tabular}{l@{\hspace{3mm}}r@{\hspace{3mm}}r}
\hline \noalign{\smallskip}
Quantity                              & Model~A          &  Model~B   \\
\noalign{\smallskip} \hline \noalign{\smallskip}
\multicolumn{3}{l}{\it (1) System parameters: }\\[1ex]
White dwarf mass $M_{\rm 1}$ (\msun) & 1.18~  & 0.93~ \\
White dwarf radius $R_{\rm 1}$ (cm)  & $\ten{4.17}{8~}$  & $\ten{6.42}{8~}$\\
Inclination $i$                       & $27.6^\circ$     & $24.4^\circ$\\
Separation $a$ (cm)                   & $\ten{9.21}{10}$ & $\ten{8.69}{10}$ \\
Corotation radius $r_{\rm co}$ (cm)   & $\ten{1.30}{10}$ & $\ten{1.20}{10}$ \\
Inner disk radius $r_{\rm i}$ (cm)    & $\ten{8.00}{8~}$  & $\ten{1.00}{10}$ \\
Outer disk radius $r_{\rm o}$ (cm), ($0.90\,r_{\rm Roche,wd}$) & $\ten{3.48}{10}$ & $\ten{3.21}{10}$ \\[1ex]
\multicolumn{3}{l}{\it (2) Temperatures:}\\[1ex]
Un-irradiated disk: \tmax\ (K)        & 58030            & (6534)\\
\hspace{24mm} \tmin\ (K)              &  6744            & (4556)\\
Irradiated disk:\hspace{4.5mm} \tirr\ (K) &              & 11000~ \\
'Reprocessed' component (K)           &                  & 40000~ \\[1ex]
\multicolumn{3}{l}{\it (3) Accretion rate and accretion luminosity for
$\alpha = 1$:\,$^1$ }\\[1ex]
X-ray accretion rate $\dot M_{\rm x}$ (g\,s$^{-1})$ & $\ten{1.24}{17\,~}$ & $\ten{1.24}{17}$ \\
Blackbody accretion rate $\dot M_{\rm bb-disk}$ (g\,s$^{-1})$ & ($\ten{3.1}{17}$) &  \\
Total accretion luminosity $L_{\rm acc}$ (\erg) &  & $\ten{2.34}{34}$ \\
X-ray luminosity $L_{\rm x}$ (\erg) &  $\ten{2.23}{34\,~}$ & $\ten{2.23}{34}$ \\[1ex]
\multicolumn{3}{l}{\it (4) Disk and non-disk/reprocessed fluxes at the Earth:\,$^2$}\\[1ex]
Disk  $>1250$\AA\ (\ergs)             & $\ten{7.61}{-10}$& $\ten{2.70}{-10}$ \\
Disk  $912-1250$\AA\ (\ergs)          & $\ten{2.85}{-10}$& $\ten{0.02}{-10}$ \\
'Reprocessed' $>1250$\AA\ (\ergs)     &                  & $\ten{4.88}{-10}$ \\
'Reprocessed' $912-1250$\AA\ (\ergs)  &                  & $\ten{2.96}{-10}$ \\[1ex]
\multicolumn{3}{l}{\it (5) Disk and non-disk/reprocessed luminosities:}\\[1ex]
Disk  $>1250$\AA\ (\erg)                & $\ten{1.23}{34}$& $\ten{4.15}{33}$ \\
Disk  $912-1250$\AA\ (\erg)             & $\ten{4.59}{33}$& $\ten{0.03}{33}$ \\
'Reprocessed' $>1250$\AA\ (\erg)        &                 & $\sim\ten{8.9}{33}$ \\
'Reprocessed' $912-1250$\AA\ (\erg)     &                 & $\sim\ten{5.4}{33}$ \\
\noalign{\smallskip} \hline \noalign{\smallskip}
\end{tabular}

$^1$ {\mdotx\ and \lx\ refer to} both poles of the white
dwarf. {\lacc\ = G\mwd$\dot M$/\rwd.} $^2$ Fluxes corrected for
interstellar extinction with $A_{\rm v}=0.47$.
\end{table}

\subsubsection{Model B: truncated disk and reverberation of X-rays}

{The alternative model~B assumes a disk, which is truncated at
$r_{\rm i} = 0.83\,r_{\rm co} = \ten{1.0}{10}$\,cm. Since the
truncated disk is intrinsically faint, we assume that heating by
irradiation raises its flux to the observed level, which requires
\tirr\ = 11000\,K. The truncated disk is represented in Fig.\,2 by an
11000\,K blackbody (short dashed curve). In Table~4, section~(4),
lines\,1 and 2, we list its UV/optical flux. The contribution from the
secondary star is minute in comparison (dotted curve for a 3100\,K
blackbody adjusted to K=15.1). Heating the disk to 11000\,K requires
that it is moderately inflated}. For a point source at the white
dwarf, a solid angle of $\Omega \simeq 2$\,sr or a {half opening
angle $\gamma =9^\circ$ is needed. If the disk is irradiated by the
funnel emission, $\gamma$ could be smaller.} In this model, much of
the observed UV radiation must be due to the reverberation of X-rays.

That the observed ultraviolet radiation is not {entirely} of disk
origin was indicated already by Mouchet's (1983) observation that the
UV flux varied by 23\% over 1.5\,h while the V magnitude stayed
constant. Furthermore, Bonnet-Bidaud et al. (1982, see also Welsh \&
Martell 1996) showed that the pulsed component has a steeper spectrum
than the mean light. When corrected for extinction, it is almost as
steep as a Rayleigh-Jeans spectrum, $f_{\nu} \propto
\lambda^{-1.9}$. The wavelength dependence of this component is
depicted by the open triangles in Fig.\,2, adjusted to 30\% of the
{total} mean flux in the $B$-band. This percentage is at the upper
end of the observed range of amplitudes (Steiner et al. 1981,
Bonnet-Bidaud et al. 1982, King \& Williams 1983, Warner \& Cropper
1984) {and is a reasonable choice if the observed variations in
amplitude represent}  variations in the visibility of the modulation
rather than true variations of the {pulsed flux}. For illustrative
purposes, we have adjusted a 40000\,K blackbody {to fit these
fluxes (Fig.\,2, long dashed curve) and the sum of the two components
to fit the total observed flux. The `reprocessed' flux is listed in
Table~4, section~(4), lines\,3 and 4.}  We do not contend that the
794\,s pulsed component {extends into the UV as indicated by the
40000\,K blackbody}, but suggest that there may exist {more than
one component of reprocessed light. Combined} they may account for
much of the UV flux. {In view of Penning's (1985) discovery of
radial velocity variations at the spin period of the white dwarf, we
suspect that a spin-modulated component may still be
hidden in the UV. The heated pole cap of the white dwarf is a viable
contender. The fact that such a component has not been found at
optical wavelengths may be due to a small amplitude of modulation
and/or} a contrived geometry (Hellier, 2003). The {sum of
truncated} disk and 'reprocessed' component is seen to fit the data as
well as the model {of a luminous disk. The present model~B has the
advantage of being internally consistent with the spin down of the
white dwarf and with the X-ray derived accretion rate \mdotx. Its
parameters depend only weakly on the choice of $M_2$.}

Is the generation of much of the UV flux by reverberation of X-rays
energetically feasible?  
{In order to test this hypothesis, we convert the components of
the observed flux $F$ in Table~4, section\,(4), to luminosities
\mbox{$L = g\,d^2F$} using appropriate (i.e. non-blackbody) geometry
factors $g$. For the `disk' flux we assume a grey limb darkening law,
$F_{\rm i} \propto (1+1.5\,{\rm cos}\,i)\,{\rm cos}\,i$, which yields
$g=1.85\,\pi$ for $i=24.3^\circ$. For the `reprocessed' component, $g$
is similar to that of the disk if the heated polar cap is the main
source, while an estimated $g\simeq 2.6\,\pi$ applies to the toroidal
geometry of the magnetically guided flow inside \ri. We use an average
$g = 2.2\,\pi$. The resulting luminosities are listed in Table~4,
section\,(5). After corrections for the contributions by the
un-illuminated disk and the secondary star, we obtain an upper limit
to the reprocessed luminosity $L_{\rm
rep,max}(>912\,\AA)=\ten{1.78}{34}$\,\erg\ = 0.80\,\lx. The
uncertainty in this number is fairly large because of possible errors
in the geometry factors and the remaining uncertainty in the X-ray
luminosity. Long-term variability of V1223\,Sgr could affect the
result, but is probably of minor importance because the AAVSO records
(Mattei 2003) show V1223\,Sgr at $V=13.1$ at the time of the ASCA
X-ray observations, in agreement with its long-term mean magnitude.
The load on the X-ray source is reduced if part of the optical
emission is of cyclotron origin, represented by the quantity $\alpha$
introduced in Sect.\,3.4, which depends on the field strength and the
pre-shock mass flow density $\dot m$ (in \gcs). From \mdotx\ and an
estimate of the area of the ring-shaped accretion region on the white
dwarf we expect $\dot m \simeq 3-10$\,\gcs. For $B \simeq 10-15$\, MG,
we then estimate $\alpha \simeq 0.05-0.15$ using the
radiation-hydrodynamic calculations of Woelk \& Beuermann (1996, their
Fig. 9, bottom panel, lower curve). For $\alpha\simeq 0.10$, $L_{\rm
rep}$/\lx\ drops to a comfortable 0.69, which can easily be accounted
for by X-ray heating of the pole cap of the white dwarf and
reprocessing in the funnel, in the accretion disk, and the irradiated
face of the secondary star (see Beardmore et al. 2000 for an analysis
of the energy-dependent X-ray albedo from the white dwarf). For
$\alpha = 0.1\pm0.05$, the accretion rate and the total accretion
luminosity increase over the values quoted in Sect. 3.5 and in Table~4
to $\dot M = (1+\alpha)\,\dot M_{\rm x} =
\ten{(1.36\pm0.27)}{17}$\,\gs\ and $L_{\rm acc} =
1.049\,(1+\alpha)\,L_{\rm x}=\ten{(2.57\pm0.83)}{34}$\,\erg, where the
errors are from Sect. 3.5 and the numerical factor in the last
relation accounts for the energy released outside \ri. Since the field
estimate given below suggests that some cyclotron emission should be
present, we accept the latter values as our best estimates.

\subsection{Magnetic moment and field strength of the white dwarf}

The different sizes of the central holes in the accretion disks imply
different surface field strengths $B_{\rm 1}$ of the white dwarfs in
models~A and B. A rough estimate of $B_{\rm 1}$ may be obtained by
equating \ri\ with $0.52\,r_{\rm \mu,sph}$ (see above). For model~B
with $r_{\rm i}\simeq \ten{1.0}{10}$\,cm , $\mu =
\ten{2.06}{33}$\,G\,cm$^3$, and for $r_1 = \ten{6.42}{8}$\,cm
(Table~4), $B_{\rm 1} = 8$\,MG. Depending on the obliquity of the
dipole, the polar field strength is in the range of $B_{\rm p} =
10-16$\,MG.
The small inner hole of the disk with $r_{\rm i}= \ten{8.0}{8}$\,cm in
model~A implies a much smaller surface field strength of about
0.5\,MG.  Our favored model~B places V1223\,Sgr at the lower range of
field strengths observed in polars. Since low-field polars are
characterized by weak soft X-ray emission, the lack of a strong soft
X-ray source in V1223\,Sgr is not surprising.}

\section{Conclusion}

We have presented an accurate parallax of V1223 Sgr which allows us to
derive the luminosities in the different wavelength bands. Based on
this result, we have tested the hypothesis that much of the UV/optical
emission is produced by the reverberation of X-rays and not by the
release of {gravitational} energy in a luminous disk. {Our
analysis is based on the assumption that we see only the X-rays from
one pole. This assumption and the high X-ray temperature reported by
Beardmore et al. (2000) lead to a total X-ray luminosity $L_{\rm x} =
\ten{(2.2\pm0.7)}{34}$, which is much higher than previously thought
and can power a large fraction of the observed UV/optical radiation by
the reprocessing of X-rays. Interestingly, this is possible without
the presence of a substantial source of XUV radiation.  Likely
reprocessing sites are the pole cap of the white dwarf, the
magnetically guided accretion flow, the disk, and the irradiated face
of the secondary star. While the latter two sites may be responsible
for the observed flux modulated at the sideband frequency (Steiner et
al. 1981), reprocessing in the former locations may produce a so far
undiscovered component, which is photometrically modulated at the spin
period of the white dwarf. The accretion rate derived from the X-ray
luminosity and X-ray temperature is $\dot M =
\ten{(1.4\pm0.3)}{17}$\,\gs, which includes a small correction for the
contribution by cyclotron emission. This result is independent of the
white dwarf mass.

The observed UV/optical/IR spectral energy distribution can equally
well be fitted by a truncated disk plus reprocessed component (model~B)
and by a luminous disk (model~A). The observed spin-down of the
white dwarf (Jablonski \& Steiner 1987, van Amerongen et al. 1987),
however, requires that the the inner edge of the disk is not too far
inside the corotation radius, a condition which is met only by the
truncated-disk model~B.

Strong internal absorption and the reprocessing of a major fraction of
the emitted X-rays is a general feature observed also in other
intermediate polars, some of which} are even more strongly internally
absorbed than V1223\,Sgr (e.g. Norton \& Watson 1989).  Simultaneous
X-ray/UV studies could shed light on the physical processes acting in
these systems. It is surprising that no such study is yet available.

\begin{acknowledgements} 
One of (KB) thanks Andrew Beardmore for providing his model fits to
the X-ray spectrum of V1223 Sgr and Coel Hellier for a stimulating
discussion on the physics of intermediate polars. {We thank the
referee John Thorstensen for pointing out errors in Table~2 and for
very helpful comments which led to an improved presentation.} This
research was supported in Germany by DLR/BMFT grant
50\,OR\,99\,03\,1. In the UK, BTG was supported by a PPARC Advanced
Fellowship. In the Unites States, partial support for TEH, BEM, and
GFB for proposal \#9230 was provided by NASA through a grant from the
Space Telescope Science Institute, which is operated by the
Association of Universities for Research in Astronomy, Inc., under
NASA contract NAS 5-26555.  This research has made use of the NASA/
IPAC Infrared Science Archive, which is operated by the Jet Propulsion
Laboratory, California Institute of Technology, under contract with
the National Aeronautics and Space Administration. This publication
also makes use of data products from the Two Micron All Sky Survey,
which is a joint project of the University of Massachusetts and the
Infrared Processing and Analysis Center/California Institute of
Technology, funded by the National Aeronautics and Space
Administration and the National Science Foundation. We acknowledge
with thanks the variable star observations from the AAVSO
International Database contributed by observers worldwide and used in
this research.
\end{acknowledgements}

\end{document}